\documentclass{aa}  

\usepackage{graphicx}
\usepackage{txfonts}
\usepackage{lipsum}
\usepackage{subcaption}         
\usepackage{lscape}             
\usepackage{placeins}           
\usepackage{xcolor}                               

\begin{document}

    \title{S-PLUS Clusters And Large-scale Environments (SCALE):}
    \subtitle{II. PZWav vs redMaPPer identification of eRosita groups}

   \author{L. Doubrawa\inst{1,2}
        \and A. Finoguenov\inst{1}
        \and E. S. Cypriano\inst{2}
        \and C. Mendes de Oliveira\inst{2}
        \and E. V. Lima\inst{2}
        \and L. Nakazono \inst{3}
        \and G. Souza \inst{1}
        \and J. Comparat\inst{4}
        \and A. Gonzalez\inst{5}
        \and R. Demarco \inst{6}
        \and A. Kanaan\inst{7}
        \and T. Ribeiro\inst{8}
        \and W. Schoenell\inst{9}
        }

   \institute{Department of Physics, University of Helsinki, P.O. Box 64, FI-00014 Helsinki, Finland
    \email{lia.doubrawaq@usp.br}
        \and Departamento de Astronomia, Instituto de Astronomia, Geof\'isica e Ci\^{e}ncias Atmosf\'ericas da Universidade de São Paulo, Cidade Universitaria, 05508-900, São Paulo, SP, Brazil. 
        \and Observatório Nacional - MCTI, Rua Gal. José Cristino 77,  Rio de Janeiro, 20921-400, Brazil
        \and  Univ. Grenoble Alpes, CNRS, Grenoble INP, LPSC-IN2P3, 53, Avenue des Martyrs, 38000, Grenoble, France
        \and Department of Astronomy, University of Florida, Gainesville, FL 32611-2055, USA
        \and Institute of Astrophysics, Facultad de Ciencias Exactas, Universidad Andrés Bello, Sede Concepción, Talcahuano, Chile  
        \and Departamento de Física, Universidade Federal de Santa Catarina, Florianópolis, SC 88040-900, Brazil 
        \and Rubin Observatory Project Office, 950 N. Cherry Ave, Tucson 85719, USA
        \and The Observatories of the Carnegie Institution for Science, 813 Santa Barbara St, Pasadena, CA 91101, USA
        }

   \date{}

\abstract{We present the construction and characterization of a multi-wavelength catalog of galaxy groups and clusters by matching optical detections from the Southern Photometric Local Universe Survey (S-PLUS) with extended X-ray emission from the first eROSITA all-sky survey data release (eRASS1). We employ a probabilistic matching framework, based on the modified Hausdorff distance, to associate galaxy systems identified by the PZWav cluster finder and characterized by the AME membership estimator with X-ray surface brightness contours. This method explicitly accounts for the photometric redshift probability distribution of galaxies and allows us to explore the critical trade-off between catalog completeness and purity. We investigate how the matched sample changes with different optical selection depths, defined by absolute magnitude cuts of $M_r < -18.5$, $-19$, $-19.5$, and $-20$ sampling redshifts within $0.08<z<0.25$, and across purity levels of 80\%, 90\%, and 95\%. We find that fainter optical cuts enhance the recovery of low-mass, low-luminosity groups, while brighter cuts favor more massive clusters and increase the effective survey volume at higher redshifts. Stricter purity requirements reduce contamination but systematically lower completeness, particularly for low-luminosity systems. The derived X-ray luminosity functions agree well with previous determinations, and the $\log N$–$\log S$ distributions confirm the high recovery rate of luminous clusters. Comparisons with the redMaPPer cluster catalog validate our approach, showing consistent trends and significant overlap, while our method offers improved completeness at the group scale. This work demonstrates a robust, flexible methodology for creating reliable multi-wavelength cluster catalogs, essential for cosmological studies and investigations of galaxy evolution in dense environments.}


   \maketitle

\nolinenumbers
\section{Introduction} \label{sec:introduction} 

Virialized dark matter halos are fundamental to both cosmology and galaxy evolution studies. Their abundance provides direct constraints on cosmological parameters, while their baryonic content reflects the efficiency of feedback processes, particularly from active galactic nuclei (AGN). Galaxy groups are ideal laboratories for resolving the role of AGN feedback due to their characteristic scale: stellar feedback processes (e.g., galactic winds) are generally insufficient to alter the properties of massive halos, whereas AGN feedback emerges as the dominant mechanism regulating star formation \citep[see][for a review]{Eckert2021}. Crucially, the binding energy of groups is comparable to the typical energy output of AGN, imprinting clear observational signatures on the warm gaseous halo.

The reliable identification of galaxy groups places stringent requirements on the depth and quality of observational data. One of the largest existing group catalogs has been constructed from the SDSS spectroscopic survey, utilizing volume-limited samples reaching absolute magnitudes of $R = -18$ at $z < 0.04$ and $R = -20.5$ at $z < 0.09$. These have been cross-matched with X-ray sources detected in the ROSAT All-Sky Survey \citep{Kosowski2025}.

The observational landscape has been transformed by the advent of the eROSITA all-sky survey, which provides X-ray data with dramatically improved sensitivity and angular resolution. In parallel, deep multi-band optical catalogs from the Southern Photometric Local Universe Survey \citep[S-PLUS,][]{Mendes2019} offer precise photometric redshifts \citep{Lima2022}. The favorable overlap between the $\sim 5,000$ square degrees S-PLUS footprint -- observed in a 12-band system with the T80-South telescope -- and the deepest portions of the first eROSITA public data release \citep[eRASS1,][]{Merloni2024} creates a powerful, multi-wavelength dataset. This synergy enables detailed investigations across multiple astrophysical pathways, including the detection and characterization of clusters in optical and X-ray bands, studies of the intracluster medium, intracluster light, and analyses of galaxy morphology and evolution.

Previous studies, such as \cite{Werner2022} and \cite{Doubrawa2024}, have demonstrated the effectiveness of the adaptive wavelet filtering technique \textsc{PZWav} \citep{Gonzalez14, Euclid2019} for cluster finding in wide-field photometric surveys. This algorithm operates directly on the photometric redshift probability density function (PDF) of galaxies, producing cluster catalogs with precise redshift estimates. Building upon this methodology, \cite{Doubrawa2023} introduced AME, an adaptive membership estimator, which uses photometric properties to assign galaxies a probability of being associated with a structure. AME further derives integrated cluster properties such as richness, optical luminosity, and total stellar mass.

These efforts are part of the S-PLUS Clusters And Large-scale Environments (SCALE) program\footnote{\url{https://splus-scale.github.io}}, which aims to investigate galaxy groups and clusters across a range of environments by combining homogeneous S-PLUS imaging with spectroscopy and multiwavelength data. In the first paper of the series \citep[][Paper I]{paperI}, we analyzed 83 previously known galaxy groups and clusters selected from the literature, deriving dynamical properties such as masses, radii, velocity dispersions, and spectroscopic membership, and presenting a detailed case study of Abell 4038.

In this paper, we extend the approach by constructing a complementary sample of systems associated with X-ray emission detected by eROSITA. We introduce the eSCALE catalog, produced through an optimized matching between clusters detected with the PZWav algorithm in S-PLUS and extended X-ray sources from eROSITA. We evaluate the purity and completeness of the resulting catalog, explore the trade-offs inherent to the matching procedure, and compare the identified systems with clusters detected using the red-sequence Matched-filter Probabilistic Percolation cluster-finding algorithm \citep[redMaPPer,][]{Rykoff2014}. This catalog provides a multiwavelength extension of the SCALE program and will enable future studies of cluster structure, galaxy populations, and environmental effects.

This paper is structured as follows. Section \ref{sec:data} describes the eRASS1 and S-PLUS datasets, the galaxy catalog construction, and the initial cluster sample selection. Section \ref{sec:methods} details our matching technique between X-ray surface brightness contours and optically selected cluster member galaxies. Our main results are presented in Section \ref{sec:results}, including the general properties of the final catalog, the X-ray luminosity function for different purity levels and absolute magnitude cuts, and detailed comparisons with the redMaPPer cluster catalogs. We summarize and discuss our conclusions in Section \ref{sec:conclusions}.
We adopt a flat $\Lambda$CDM cosmology, with $H_{0}=70$ km\,s$^{-1}$\,Mpc$^{-1}$, $\Omega_{m}=0.3$. Magnitudes are given in the AB system.

\section{Data} \label{sec:data}

\subsection{eROSITA survey}

We utilize X-ray data from the first public data release of the eROSITA all-sky survey \citep[eRASS1,][]{Merloni2024}, which was performed with the eROSITA instrument aboard the Spectrum–Roentgen–Gamma (SRG) mission. The eRASS1 dataset comprises overlapping sky tiles, each with an effective area of approximately 10 square degrees, ensuring contiguous coverage over extensive sky regions.

Our analysis targets the soft X-ray emission originating from galaxy groups and clusters. Previous studies have highlighted the presence of significant large-scale diffuse emission in the softest eROSITA bands, including emission associated with cosmic filaments. This diffuse emission complicates source detection, especially on large angular scales, due to source confusion. Furthermore, strong foreground emission from the Milky Way in the 0.1--0.6 keV band substantially diminishes the sensitivity to extragalactic sources in terms of detectable source counts.

To alleviate these issues, we confine our analysis to the 0.6--2.3 keV energy band. This band is relatively unaffected by Galactic foreground emission, reduces contamination from large-scale diffuse structures, and enhances the detectability of extended X-ray emission from galaxy clusters. Additionally, the 0.6--2.3 keV band allows for direct comparison with detections from the ROSAT All-Sky Survey \citep[RASS,][]{ROSAT1993} and circumvents potential contamination from solar X-ray leakage, which is known to affect eROSITA data at lower energies.

\subsection{S-PLUS survey}

The Southern Photometric Local Universe Survey \citep[S-PLUS,][]{Mendes2019} is a photometric survey designed to cover 9,000 square degrees of the southern sky using a 12-filter system, comprising five broad-band and seven narrow-band filters. Observations are conducted with the 80 cm T80-South telescope at the Cerro Tololo Inter-American Observatory (CTIO) in Chile. Each pointing covers a field of view of $1.4 \times 1.4$ square degrees, resulting in an imaged area of $\sim2$ square degrees per field. The combination of narrow-band photometry and accurate photometric redshifts makes S-PLUS particularly well suited for identifying and characterizing galaxy clusters and groups over wide areas.

The fifth data release \citep[DR5][]{dr5} of S-PLUS encompasses approximately 4,600 square degrees of the sky. However, the effective overlap with the eROSITA survey is about 2,500 square degrees. According to \citet{Lima2022}, objects with $r$-band magnitudes fainter than 21.3 achieve a photometric redshift accuracy of $\sigma_{\mathrm{NMAD},z} = 0.0189$, which improves to $\sigma_{\mathrm{NMAD},z} = 0.006$ for objects with $r \leq 17.5$. These values attest to the precision and high quality of the S-PLUS photometric redshifts. To ensure a complete galaxy catalog, we limit our analysis to galaxies brighter than $r_{auto} = 21$. The magnitudes used in this work are derived from Kron apertures and are corrected for Galactic extinction following the S-PLUS standard calibration pipeline \citep{Schwarz2025A}.

To minimize contamination from stars and spurious sources, we cross-match the S-PLUS data with the spectroscopic compilation of southern hemisphere objects presented by \citet{Erikspecz2024}, retaining only sources classified as ``GALAXY''. For objects without a spectroscopic counterpart, we select those classified as galaxies using a machine-learning approach specifically calibrated for S-PLUS \citep{Nakazono2021}. This combined procedure ensures a high-purity galaxy sample across the entire magnitude range. We further remove regions affected by bright stars and imaging artifacts by applying the SExtractor software \citep{Bertin1996} flag \texttt{sex\_flag\_det < 3}. Additional details on calibration and survey specifications are provided in the data release documentation\footnote{https://splus.cloud/documentation/dr5}.

Although the full S-PLUS photometric redshift coverage spans $0 < z < 0.5$ \citep[see][]{Mendes2019}, we restrict our sample to the range $0.08 < z < 0.25$. The low-z limit is motivated by the presence of the prominent low-redshift large-scale structure, the Horologium–Reticulum supercluster, within the footprint, which requires special consideration. At $z > 0.25$, the eROSITA data are only deep enough to detect galaxy clusters, which fall outside the main focus of this paper.

\subsubsection{Optical galaxy cluster and group detection}

The S-PLUS cluster and group catalog (Doubrawa et al., in prep.) was constructed using the cluster-finder algorithm \textsc{PZWav}. Within the overlapping area between S-PLUS and eROSITA, we identify 49,099 cluster and group candidates in the redshift range $0.08 < z < 0.25$.

\textsc{PZWav} is an overdensity-based algorithm that detects galaxy systems by generating two-dimensional density maps across a series of redshift slices of fixed width $\Delta z = 0.01$. Each density map is created using the sky positions of galaxies and by integrating their photometric redshift probability distribution functions (PDF) over the redshift limits of the slice. A cluster or group candidate is identified as a peak in the density map with a signal-to-noise ratio exceeding a threshold of $\mathrm{S/N}_{\mathrm{thr}} = 4$. The noise level is estimated as the standard deviation of a Gaussian approximation derived from density maps constructed using randomized galaxy sky positions.

The center of each detected structure is defined as the location of the highest local peak within a cylindrical volume of radius $dr_{\mathrm{lim}} = 1,500$ kpc and redshift interval $dz_m = 0.03$, a criterion adopted to prevent multiple detections of nearby substructures. The cluster redshift is then determined as the median photometric redshift of galaxies within a redshift interval of $\Delta z = 0.02$ (twice the redshift bin width) around the candidate peak.

\subsubsection{Galaxy membership with AME}

To further characterize the detected systems, we employ the Adaptive Membership Estimator (AME) algorithm \citep{Doubrawa2023}. AME combines the Hierarchical Density-Based Spatial Clustering of Applications with Noise (HDBSCAN) algorithm \citep{Campello2014} with galaxies' photometric redshift PDF to estimate membership probabilities.

Briefly, AME selects galaxies within an aperture radius of 1.5 Mpc centered on the cluster sky position and satisfying ${|z_{\mathrm{cl}} - z_{\mathrm{gal}}| < 3\,\sigma_{\mathrm{NMAD},z}}$, where $z_{\mathrm{gal}}$ is the galaxy photometric redshift. For each selected galaxy, a redshift is randomly drawn from its PDF. To reduce contamination from field galaxies, a $3\,\sigma$ clipping is applied to the overall velocity distribution. HDBSCAN is then applied to the remaining galaxy sample to identify spatially connected structures. Galaxies linked by HDBSCAN are classified as cluster members, while isolated galaxies are flagged as interlopers. This procedure is repeated 100 times, and the membership probability for a galaxy, $P_{\mathrm{mem}}$, is defined as the fraction of iterations in which it is identified as a cluster member.

Cluster properties are derived as probability-weighted quantities. The richness, $\lambda$, is defined as the sum of the membership probabilities of all galaxies associated with a cluster:
\begin{equation}
\lambda = \sum_i P_{\mathrm{mem},i}.
\end{equation}

To investigate the dependence of cluster properties on optical depth, we construct four subsamples with different richness estimates defined by the absolute magnitude limits: $M_r < -18.5$ (volume-limited up to $z = 0.13$), $M_r < -19$ ($z = 0.15$), $M_r < -19.5$ ($z = 0.2$), and $M_r < -20$ (extending to $z = 0.25$). Absolute magnitudes are computed from apparent magnitudes using,
\begin{equation}
M_r = r - 5\log_{10}(d_{\mathrm{Mpc}}) - 25,
\end{equation}
where $r$ is the apparent $r$-band magnitude and $d_{\mathrm{Mpc}}$ is the luminosity distance in megaparsecs.

\section{eROSITA source detection and identification} \label{sec:methods}

\begin{figure*}
    \centering
    \includegraphics[width=0.4\linewidth]{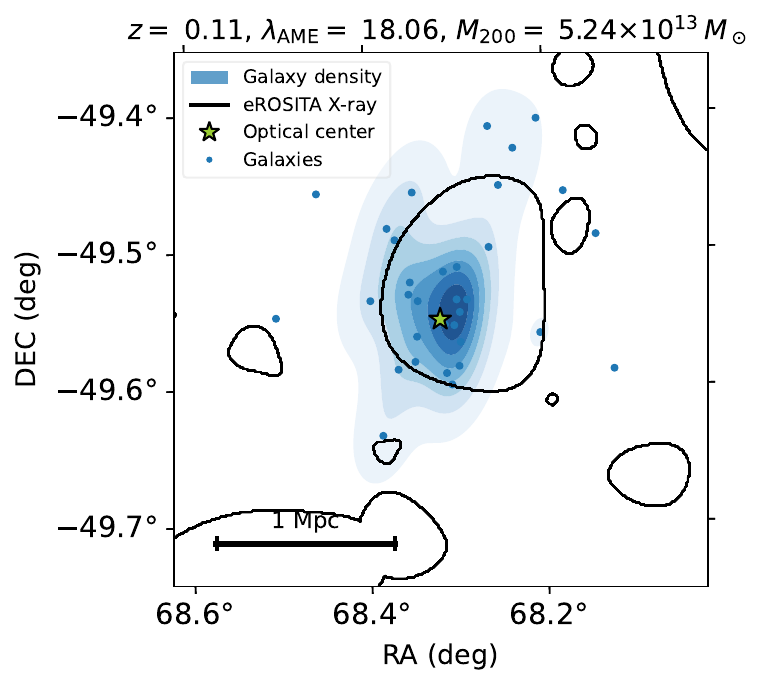}
    \includegraphics[width=0.4\linewidth]{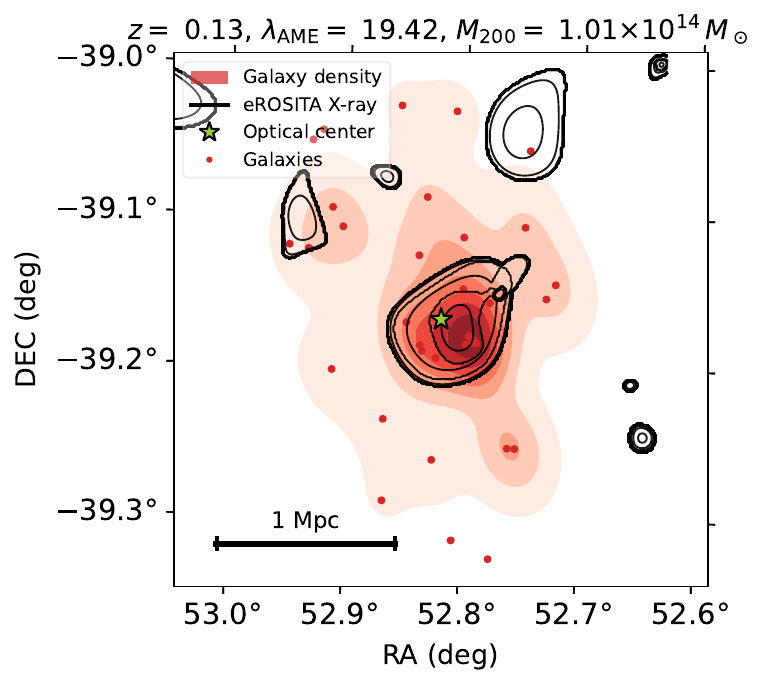}
    \includegraphics[width=0.4\linewidth]{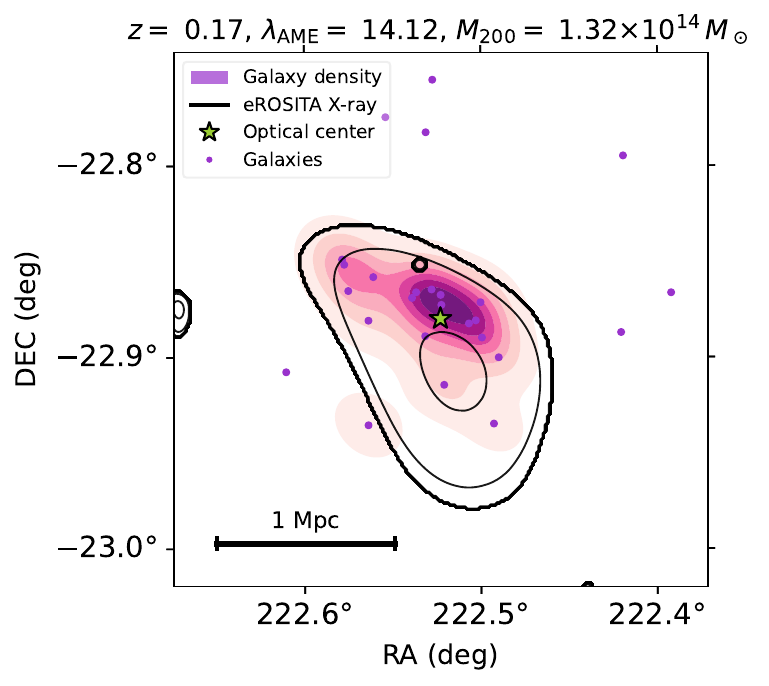}
    \includegraphics[width=0.4\linewidth]{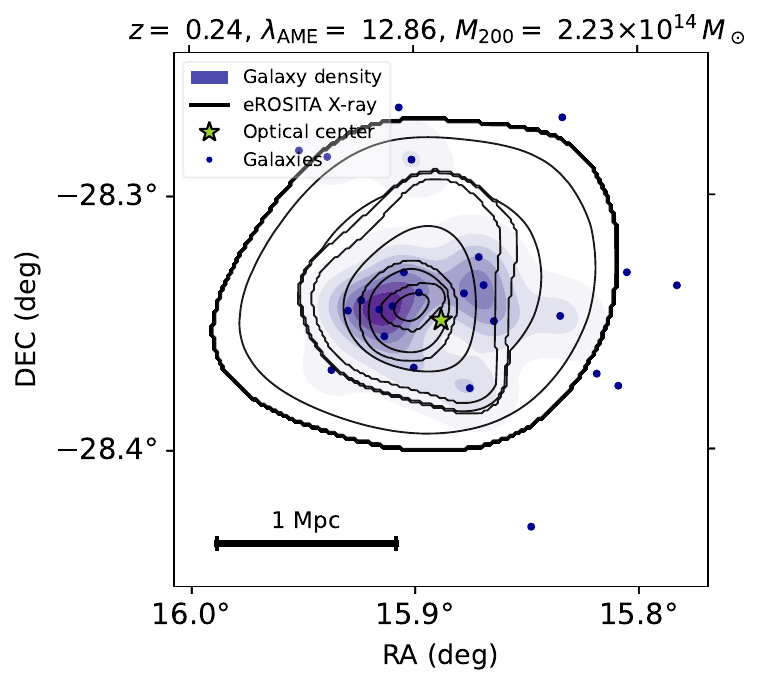}
    \caption{Examples of detected clusters, along with their associated member galaxies. Colors indicate the adopted absolute magnitude cut: $M_r < -18.5$ as light blue; $M_r < -19$ as red; $M_r < -19.5$ as purple; and $M_r < -20$ as dark blue. The same color code is used throughout the paper. Shaded regions show overdensities weighted by the galaxy membership probability. Gray contours indicate the eROSITA X-ray emission. The optical center does not necessarily coincide with the X-ray center.}
    \label{fig:examples}
\end{figure*}

We utilize the first public data release of the eROSITA-DE survey (eRASS1), which consists of overlapping sky tiles, each covering approximately 10 square degrees. 
Given that most of the eRASS1 data consists of single-pass observations, data screening based on background levels does not significantly improve source detection. Instead, we adopt a reproducible approach that uses the released datasets without modification, i.e., the public eROSITA images, exposure, and background maps.
We perform a wavelet analysis of the eROSITA tiles on apstial scales of 1/8 to 16 arcminutes using the released exposure and background maps.
The public background maps are constructed by smoothing the residual emission after source detection and are not suitable for studying extended emission on scales comparable to the smoothing kernel (approximately half a degree). 
Additionally, artifacts near bright sources required us to exclude sky regions adjacent to 20 bright sources and to reiterate the background estimation around them.

The eROSITA survey-averaged point spread function (PSF) is 0.5 arcminutes, while the scales of interest for this study range from 2 to 4 arcminutes. When extracting a source's flux, we remove areas corresponding to sources detected on spatial scales of 0.5 arcminutes or less. 
In this analysis, we excise the emission areas detected on scales up to 0.5' in the 0.2-2.3 keV band. These are 664 thousand unique source areas (counting blended sources within an area as one), out of which 566 thousand areas contain the sources present in the catalog of \cite{Merloni2024}. From their catalog, only seven thousand sources with reported detection significance above 4$\sigma$, and showing no extent, are located outside of our areas. Five thousand of these are associated with known Supernova Remnants, stellar spikes, Magellanic clouds, as well as confusion limited zone near the Ecliptic pole (due to deeper exposure). 
These steps also effectively remove cool-core emission, similar to our analysis of deep X-ray fields, thereby allowing our weak-lensing mass calibrations to be directly applicable.

For source identification, we follow the approach developed for identifying RASS sources using SDSS groups \citep{Kosowski2025}, enabling a direct comparison of results. 
This method relies on identifying groups by matching X-ray contours with the member galaxies that define their outskirts.
A uniform baryonic overdensity threshold used in RASS analysis corresponds to an eROSITA flux of $10^{-6}$ counts s$^{-1}$ pixel$^{-1}$ in the 0.6--2.3 keV band, assuming $4^{\prime\prime}$ pixels and extragalactic hydrogen column density conditions, $N_H<5\times10^{20}$ cm$^{-2}$. Corrections up to $N_H<7\times 10^{20}$ cm$^{-2}$ (definition of extragalactic X-ray sky) are minor. We store the contours of X-ray emission obtained from wavelet decomposition on scales of $2-4^{\prime}$ and $8-16^{\prime}$. These contours are generated using the \textsc{DS9} software with a smoothing of a few pixels.

Each contour defines a flux extraction region, which we convert to a polynomial region in \textsc{DS9}. We assign a unique identifier to each contour, composed of the field ID and a sequential number within the field. Due to the overlap between eROSITA sky tiles, some sources are duplicated. These duplicates are cleaned in the final catalog production stage, after applying stringent matching criteria. Contours that are incomplete (i.e., only part of the source falls within a tile) are removed prior to identification.

In total, we detect 39 thousand sources on spatial scales of $2-4^{\prime}$ with a point-source-free flux above $2\sigma$. In cleaning for point sources, we excise the flux around detected sources on scales below 1 arcminute and, in addition, subtract the expected contribution to the zone of extended emission, based on in-flight measurements of the eROSITA PSF. 
We also provide a full flux estimate of the point sources to match the luminosity estimates used in the literature. In the overlap area between eRASS1 and S-PLUS, we have 6790 X-ray sources.

To match the two-dimensional shapes of X-ray sources and optical groups, we employ the modified Hausdorff distance, which was successfully applied to the RASS-SDSS catalog \citep{Kosowski2025}. We further develop this method to brute-force sample the matching criteria and to account for the probabilistic nature of galaxy membership. We define matching criteria using volume-limited subsamples of galaxies and the corresponding richness of the system (the sum of membership probabilities). The first criterion requires that the sum of membership probabilities of galaxies within a specified distance to the X-ray contours exceeds a given threshold. We sample the following threshold values: 2, 3, 4, 5, 6, 8, and 10. For comparison, \cite{Kosowski2025} only studied median values and a minimum number of members equal to 10. 
To assess chance associations, we generate a random catalog of groups by adding $10^{\circ}$ to the Right Ascension (modulo 360) of real groups. This method effectively conserves the intergalactic distances between sources, and has minimal enhancement of groups ($\sim0.0005$) produced by the clustering signal at the corresponding redshift range of this work.
The random galaxy catalog overlaps with the eRASS1 data by an additional 15\%, as it uses SCALE catalogs outside of the eRASS1-DE footprint. The distance to the X-ray is set to yield a given ratio of random to real identifications, which is sampled between 10\% and 95\%. We correct for the difference in the area between real and random catalogs and vary the value of the distance to X-ray emission till this criterion is satisfied. 

For the next step, we retain only member galaxies from the real and random groups that passed the matching criteria. The matching criteria involve two parameters: the fraction of contour points \citep[sampled from 10\% to 100\%, for comparison,][only considered 50\%]{Kosowski2025} and the target purity level of the catalog, sampled at 80\%, 90\%, and 95\% values. We define the purity level as one minus the ratio of random to real identifications multiplied by 100. This yields approximately 40,000 matching combinations, from which we select the combination that maximizes the number of matches at a fixed purity. We also examine the 5 best matches for consistency in the results. High purity requirements (95\% or more) are typically better satisfied by using higher summed membership probability values (typically between 3 and 5) and by requiring a larger fraction of the X-ray contour to be matched (80--100\% of contour points). The largest catalogs sample 50\% of contour points and use lowest summed membership probability values (2 or 3). Examples of matched systems can be found in Fig.\,\ref{fig:examples}.

We consider full catalogs of X-ray sources in the matching for completeness. When applying the cleaning of the flux from point source contamination and subsequent removal of sources with low significance of residual flux estimate, we noticed an improvement in the purity of identifications by 4-8\%.
Using the largest catalogs, we evaluate that these two matching steps contribute nearly equally to the final purity, with a match to X-ray contour being a more demanding criterion. 
In case of multiple matches, we report only the best match defined as having a high sum of probabilistic membership and a higher fraction of X-ray contour. 

We compute fluxes and X-ray luminosities in the rest-frame 0.1--2.4 keV band using our calibrations for the K-correction given the flux estimates in the observed 0.6--2.3 keV band, the exact hydrogen column density toward each source, an iterative temperature estimate from the $L-T$ relation of \cite{Markevitch1998}, and a correction for missing flux based on the effective radius of the contour. The resulting values are released in an online catalog available via the Centre de Données astronomiques de Strasbourg (CDS).
We estimate the total mass of the sources based on the weak lensing calibrations of the $L_x-M_{200}$ relation of \cite{Leauthaud2010}. A consistency of this scaling relation with eRASS1-based luminosities for clusters at $z<0.6$ has been demonstrated by \cite{pederneiras}. 
We repeat these calculations also for the random catalogs and subtract them when reporting statistical properties of the eSCALE groups.

In this paper, we compare the optical catalog produced by PZWav to the one obtained by the redMaPPer algorithm \citep{Rykoff2014}. The use of X-ray sources validates the true nature of the optical cluster, as the presence of X-ray emission safeguards against the projection effects. 
The ``RedMaPPer'' optical catalog was obtained by running Version 8 (Python) of the redMaPPer algorithm \citep[e.g., as in][]{ider2020cosmological, Kluge2024}, utilizing the photometric data from the 9th (Northern Hemisphere) and 10th Data Release of the DECam Legacy Survey \citep[DECaLS,][]{dey2019overview}. In order to exclude the role of matching in selecting the counterparts, we have repeated all the matching steps to create a catalog of identified sources using the redMaPPer membership catalog. The matched catalog will be addressed to as ``eRedMaPPer''. Due to the extensive volume of the RedMaPPer catalog, we restrict the eRedMaPPer sample to $z < 0.2$. More details are presented in Section\,\ref{sec:red_comp}.

\section{Results} \label{sec:results}

\subsection{General properties of the catalogs} \label{sec:Properties}

\begin{figure}
    \centering
    \includegraphics[width=\linewidth]{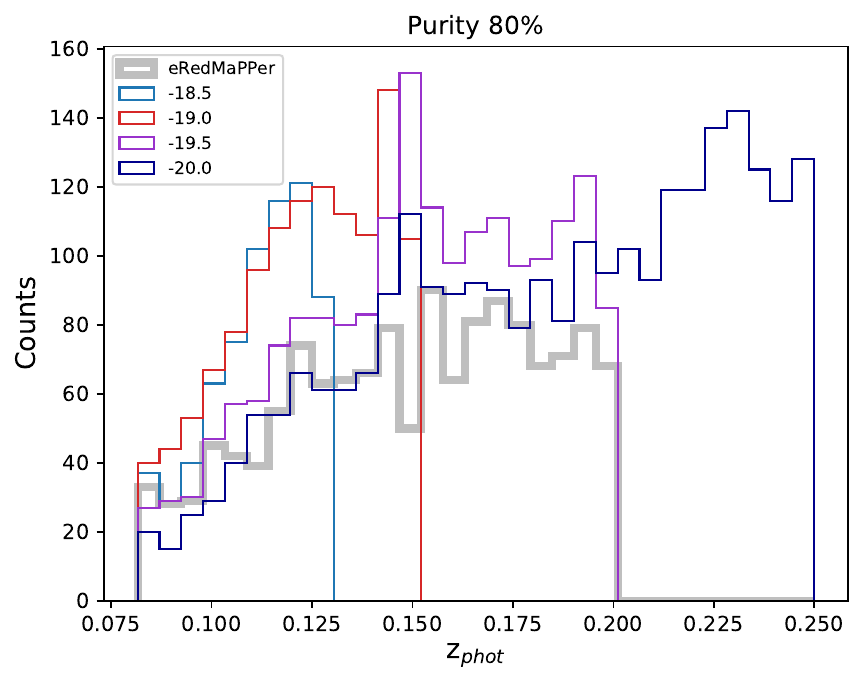}
    \includegraphics[width=\linewidth]{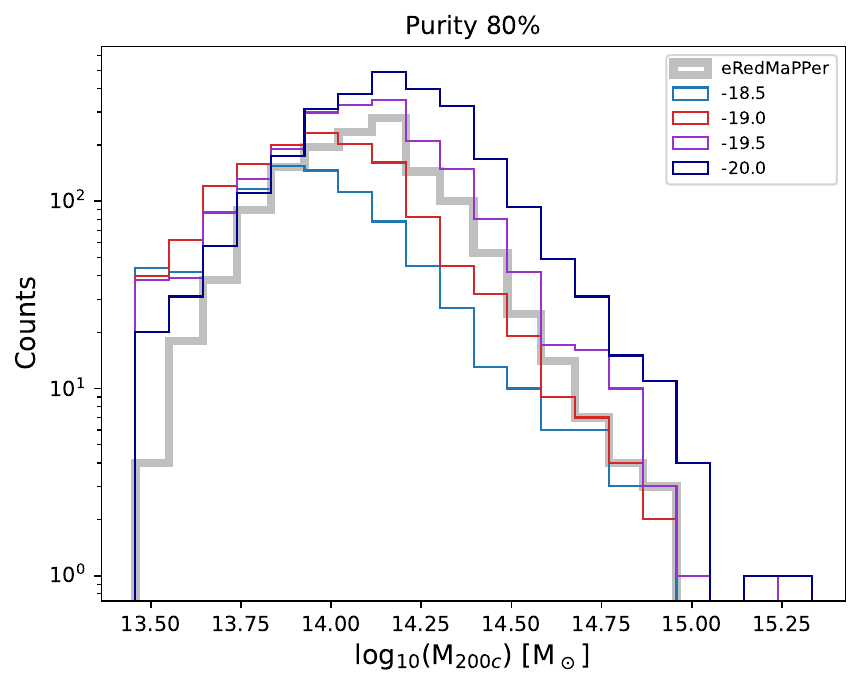}
    \caption{Top panel: Redshift distribution of the cluster catalog at a purity level of 80\%. Bottom panel: Mass distribution derived from the $L_x-M_{200}$ relation of \cite{Leauthaud2010} for the same purity. Colors indicate the adopted absolute-magnitude cut. eRedMaPPer distributions are shown for comparison in gray. There is a tradeoff associated with $M_r$, as brighter cuts reduce the raw number of galaxies but increase the fidelity of the photometric redshifts.}
    \label{fig:prop}
\end{figure}

In this work, we investigate how different optical galaxy membership definitions influence the matching between X-ray emission contours and optically detected galaxy systems, and validate our results through direct comparisons with the redMaPPer catalog. 
The optical catalog is characterized by cluster richness estimates computed using four distinct absolute-magnitude limits: $M_r < -18.5$, $M_r < -19$, $M_r < -19.5$, and $M_r < -20$. These cuts correspond to different effective volume limits and probe distinct galaxy populations used to trace the cluster outskirts.

Figure~\ref{fig:prop} presents the redshift (upper panel) and mass (lower panel) distributions of the matched systems for a fixed purity level of $P = 80\%$, shown separately for each magnitude cut. Due to the adopted redshift limits of this study, our sample is limited to low-mass groups above $2.3\times10^{13}$ M$_\odot$. Although all matched catalogs originate from the same parent optical catalog, their resulting properties differ significantly depending on the galaxy population used for matching. This highlights the sensitivity of the matching procedure to the adopted magnitude limit.

Overall, the histograms display broadly similar shapes across all magnitude selections. As the effective survey volume increases toward brighter luminosity cuts, both the total number of optically detected systems and the number of matched counterparts rise. This increase is not uniform across mass, since brighter cuts preferentially boost the identification of high-mass systems.

A similar trend is observed in the redshift distributions, where progressively brighter cuts result in a gain of the sampled volume but also show a modest reduction in the total number of matches within the overlapping redshift intervals. This behavior illustrates the trade-off introduced by stricter luminosity thresholds, which slightly reduces the sensitivity to low-mass structures.

However, at low redshift ($0.08 < z < 0.10$), the $M_r < -19$ cut recovers a number of low-mass systems that is larger than the deeper $M_r < -18.5$ selection. In this redshift interval, 71\% (155 systems) of the matched sources at $M_r < -19$ without a counterpart at $M_r < -18.5$ have masses below $M_{200c} = 10^{14}\,\mathrm{M}_\odot$, indicating an improved recovery of the group regime. While deeper cuts include fainter galaxies -- increasing the raw number counts -- they also introduce sources with larger photometric uncertainties and greater susceptibility to projection effects, especially at low redshift. These systematics reduce the contrast between genuine low-mass systems and the background. In contrast, this slightly brighter cut favors more luminous and reliably identified member galaxies, enhancing the sensitivity to low-mass structures.

One might then expect even better performance from the $M_r < -19.5$ cut. In practice, this selection reflects the transitional behavior between the regimes traced by the $-19$ and $-20$ cuts. While the total number of detections in the previously sampled redshift range is slightly reduced, the larger comoving volume allows additional matches at higher redshifts, increasing the number of more massive systems and shifting the peak of the mass distribution toward higher values.

This trend becomes even more pronounced for the $M_r < -20$ selection: the progressively brighter cuts improve robustness against projection effects, but exclude the fainter galaxy population that traces lower-mass halos. A similar behavior was reported by \citet{Doubrawa2024}, who demonstrated that group candidates identified through faint members can become undetectable when brighter luminosity thresholds are applied. Simultaneously, the larger comoving volume associated with the $M_r < -20$ cut enhances the detection of massive systems, shifting the peak of the mass distribution toward higher values by approximately 0.25 dex relative to $M_r<-19$ (see Fig.\,\ref{fig:prop}, bottom panel).

\begin{figure}
    \centering
    \includegraphics[width=\linewidth]{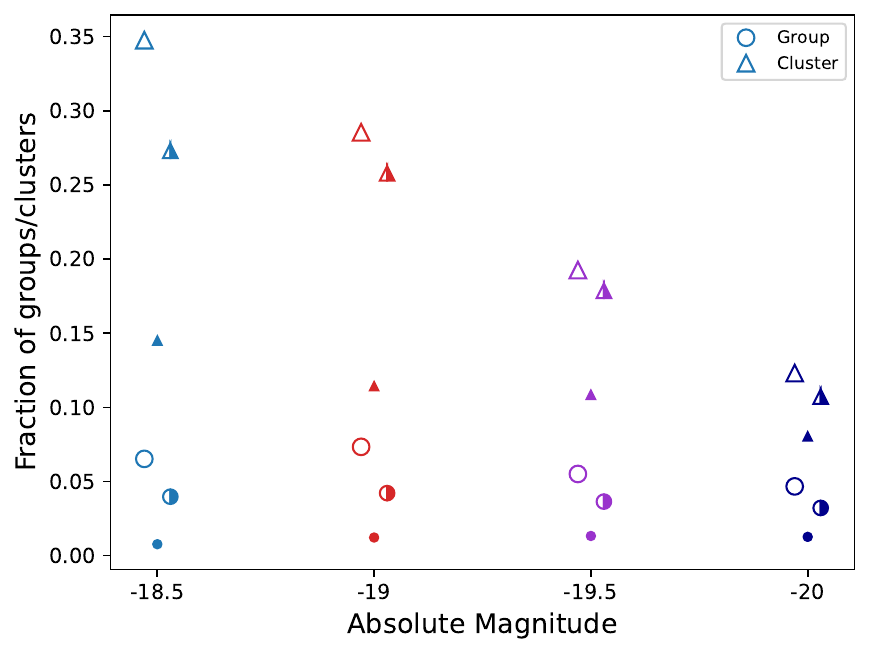}
    \caption{Fractions of clusters (triangles) and groups (circles) in each catalog for purities of 80\% (open symbols), 90\% (semi-filled), and 95\% (filled). Groups (clusters) are defined as detections with richness below (above) the threshold corresponding to $M_{200c} = 10^{14}\,{\rm M}_\odot$, derived from the mass–richness scaling relations. Colors indicate the adopted absolute-magnitude cut. While the renormalization maintains a roughly constant group fraction, the cluster fraction is reduced by the matching sensitivity towards brighter cuts.}
    \label{fig:prop2}
\end{figure}

\subsection{Group and cluster fraction}

The purity level controls the strictness of the matching criteria, with higher values imposing more stringent conditions. This choice affects the recovered population and reflects differing sensitivities to low- and high-mass systems. Figure~\ref{fig:prop2} illustrates these effects by showing how the fractions of galaxy groups and clusters, computed relative to the optical catalog, vary with purity level and absolute magnitude selection.

Since direct mass estimates are only available for the matched subsample, we use mass–richness scaling relations to define a richness-based proxy for separating galaxy groups from clusters. The resulting richness thresholds are $\lambda = 25$, 19, 15, and 10 for the $M_r < -18.5$, $-19$, $-19.5$, and $-20$ selections, respectively. These thresholds are derived from scaling relations fitted to the purest subsample (purity $= 95\%$) in order to minimize contamination. The relations are obtained via Bayesian linear regression using the \textsc{linmix} code \citep{Kelly2007}, which accounts for measurement uncertainties in both X-ray mass and optical richness.

Applying these richness thresholds, we find that the group fraction remains approximately constant across all magnitude cuts and purity levels. This behavior indicates that, although the survey volume and richness estimates vary, the relative proportion classified as groups remains stable due to the renormalization of the classification. The only exception is the $M_r < -19$ selection at $P = 80\%$, where the matching yields a slightly higher fraction of groups, increasing from 6\% to 8\%. This result is consistent with the previous discussion, indicating a modestly enhanced sensitivity to low-mass and low-richness systems.

In contrast, the cluster fraction systematically decreases as the absolute-magnitude cut becomes brighter for purity levels of 80\% and 90\%. At $P = 95\%$, however, the cluster fraction converges to a plateau of approximately 11\% across all magnitude selections. The $M_r < -18.5$ cut shows slightly higher sensitivity to high-mass systems (increasing from 11\% to 14.5\%), whereas the $M_r < -20$ cut exhibits reduced sensitivity (decreasing from 11\% to 8\%). 

This trend suggests that stricter optical selections (brighter magnitude cuts) preferentially exclude galaxies in the lower-contrast cluster outskirts, thereby reducing the recovery sensitivity. The associated increase in survey volume does not fully compensate for this loss at moderate purity levels. At the highest purity threshold, however, the effect is mitigated, resulting in a stable recovery fraction for the most massive systems.

\subsection{Flux distribution and completeness}

\begin{figure}
    \centering
    \includegraphics[width=\linewidth]{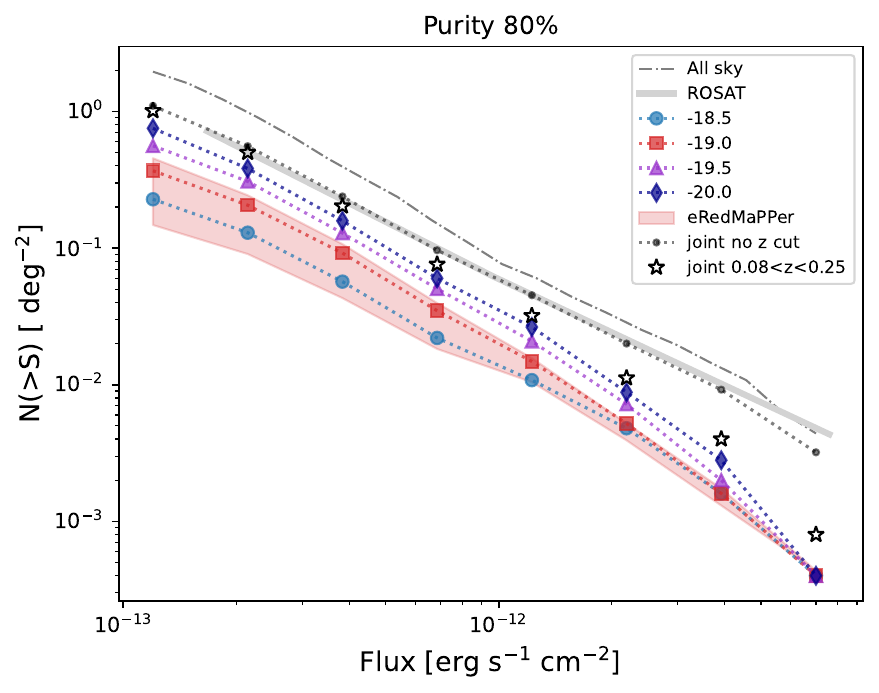}
    \caption{$\log N$–$\log S$ distribution of cluster counts as a function of X-ray flux. Symbols represent different absolute-magnitude cuts. The gray dot-dashed curve shows all X-ray sources within the S-PLUS footprint. Red shaded regions represent the eRedMaPPer catalog over the redshift range $0.08 < z < 0.2$. Star symbols indicate the combined contribution of the eSCALE and eRedMaPPer catalogs. The gray solid line shows the ROSAT All-Sky Survey relation \citep{Vikhlinin1998}, which is recovered when including the low-redshift contribution (small gray dots).}
    \label{fig:flux_dist}
\end{figure}

To further explore the completeness of the matched catalogs, we examine the number of clusters as a function of X-ray flux, i.e., the $\log N$–$\log S$ distribution, by comparing the number of matched systems to the total population of extended X-ray sources detected within the eROSITA footprint (after cleaning the X-ray sample to remove the PSF tails of bright point sources). Figure~\ref{fig:flux_dist} shows the resulting distributions for different absolute-magnitude cuts at a purity level of 80\%.

The results indicate a high recovery rate for high-flux clusters, with all curves exhibiting broadly similar shapes. As expected, the total number of detected systems increases with the effective survey volume associated with brighter absolute-magnitude limits. Between the $M_r < -18.5$ and $M_r < -19$ selections, the number of matched structures increases by a factor of $\sim1.4$. A further increase of $\sim1.5$ is observed when moving to the $M_r < -19.5$ cut, followed by an additional factor of $\sim1.3$ when applying the $M_r < -20$ selection. 
At the high-flux end, the $M_r < -20$ sample provides the most complete characterization of the bright cluster population. 

The gray shaded region shows the $\log N$–$\log S$ relation derived from the ROSAT All-Sky Survey cluster sample of \cite{Vikhlinin1998}. This catalog comprises 203 X-ray–selected clusters detected over a large solid angle and is commonly used as a benchmark for bright flux regimes. 
The $M_r < -20$ curve follows the general trend of this reference relation, indicating that the brightest optical selection successfully recovers a substantial fraction of the low-flux systems traced by the X-ray survey. However, the highest-flux end appears consistently underrepresented, which is primarily driven by the adopted low-redshift limit (we check for further effects on flux calculations in Appendix\,\ref{ap_full_lx}). This agreement also demonstrates that we have selected the most frequent spatial scales for X-ray cluster detection at those fluxes. Selected spatial scales correspond to typical cluster appearance at low-to-intermediate redshifts, sampled well by SCALE catalogs and within the eROSITA sensitivity.
The excess of extra unidentified sources at low fluxes can be ascribed to the detection of multiple point sources, each below the flux limit for the point source detection ($5 \times 10^{-13}$erg s$^{-1}$ cm$^{-2}$ in the area used for constructing logN-logS), but with the combined flux in the detected range. In the future, the catalogs can be improved using the full data release of eROSITA, which has a factor of 4 increase in the point source sensitivity and would resolve those sources.

\subsection{X-ray luminosity function} \label{sec:luminosity}

\begin{figure}
    \centering
    \includegraphics[width=\linewidth]{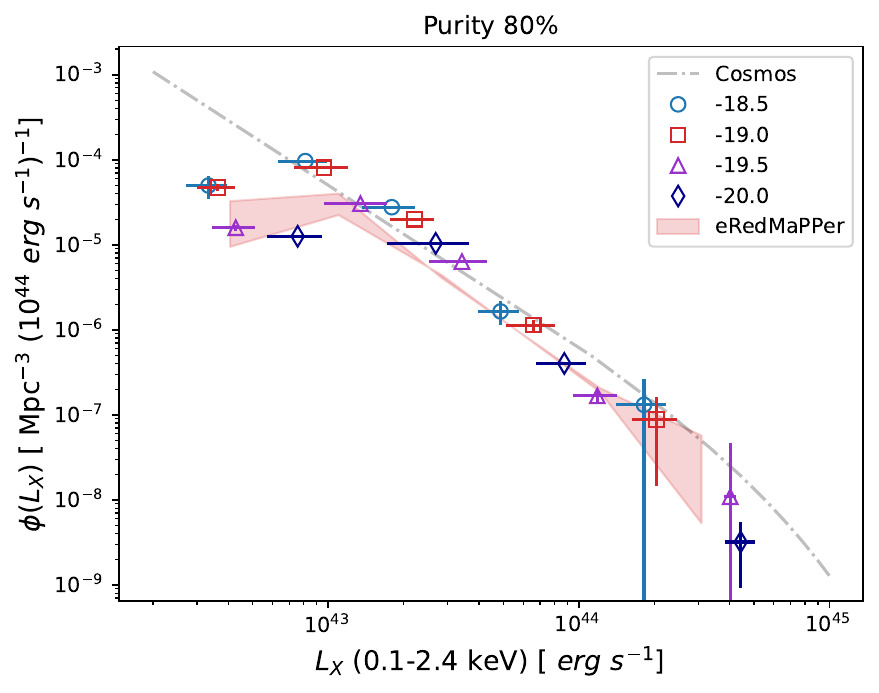}
    \includegraphics[width=\linewidth]{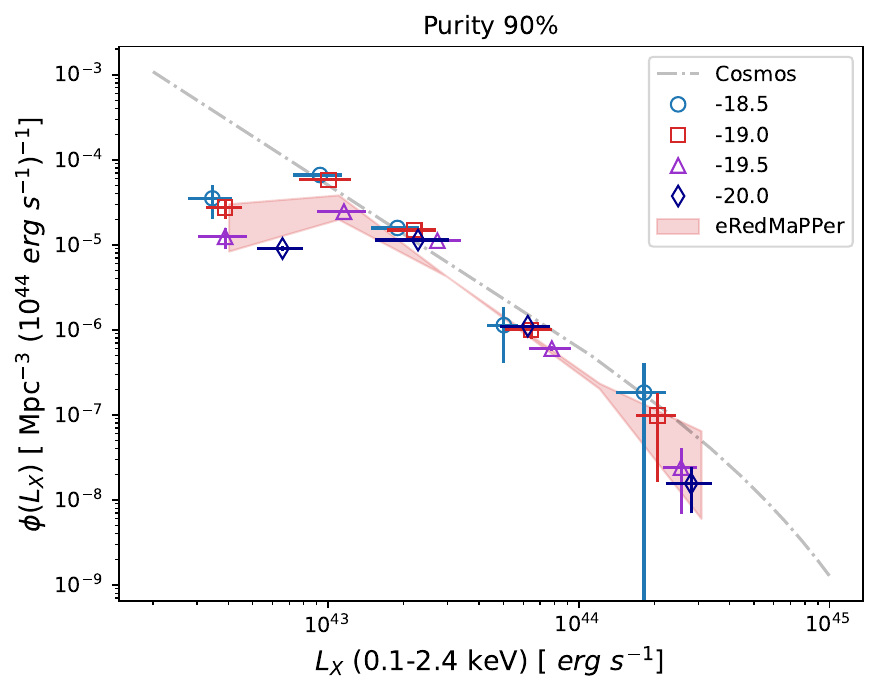}
    \includegraphics[width=\linewidth]{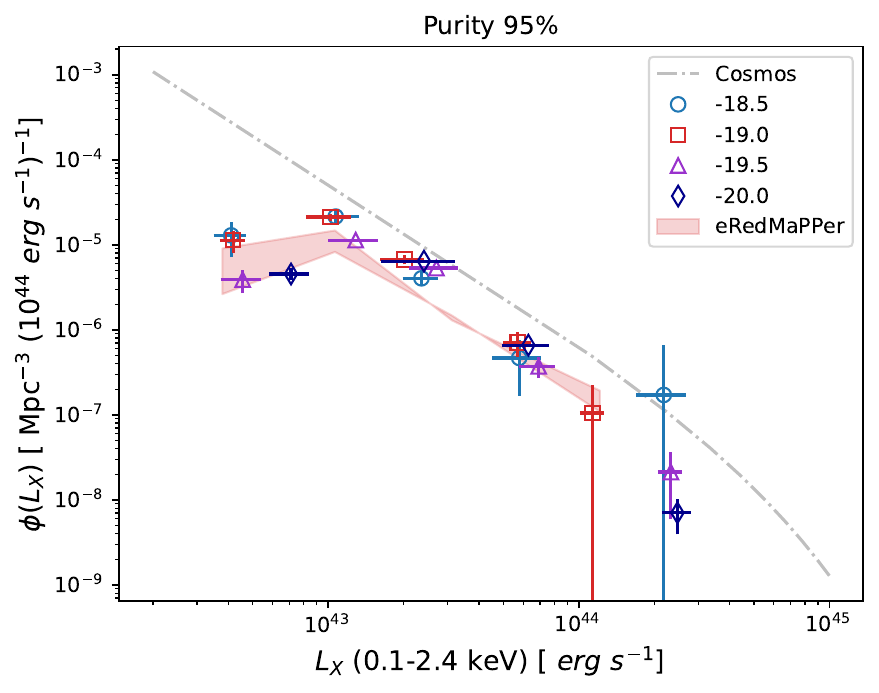}
    \caption{Cluster X-ray luminosity functions for three different purity levels. Values are corrected for contamination. Symbols represent different absolute-magnitude cuts applied to the eSCALE sample. Red shaded regions show the eRedMaPPer catalog over the redshift range $0.08 < z < 0.2$. The results show good agreement within samples and with the gray dot-dashed curve from COSMOS observations \citep{Finoguenov2007}.}
    \label{fig:XLF}
\end{figure}

X-ray luminosity functions (XLFs) provide a fundamental characterization of the galaxy cluster population, linking cluster abundance to halo mass and tracing the growth of structure in the Universe. We compute the XLF, $\phi(L_X)$, defined as the number of clusters per unit comoving volume and per unit X-ray luminosity, for the group and cluster sample at all considered purity levels.

We divide the luminosity range $42.20 < \log_{10}(L_X / \mathrm{erg\,s^{-1}}) < 44.92$ into eight bins, merging the two highest-luminosity ones to ensure a minimum of eight clusters per bin. The representative luminosity for each bin is taken as the median luminosity of the systems within it. To ensure a fair comparison with the theoretical curve, we corrected the observed XLF by weighting the number density in each luminosity bin by the corresponding catalog purity rate. 

Figure~\ref{fig:XLF} presents the resulting XLFs for the different absolute-magnitude cuts and purity selections. The error bars are calculated as the Poisson uncertainty divided by the number of sources within each bin. For reference, we compare our measurements with the XLF derived by \cite{Finoguenov2007} from deep X-ray observations in the COSMOS field, which were found to compare well to the nearby RASS data. 

Overall, we find good agreement between our measurements and the literature XLF. We observe a consistent excess of roughly a factor of 1.7 in the number of systems within the $1\times10^{43}$ erg s$^{-1}$ relative to the reference curve (Cosmos). We check for possible random associations in Appendix\,\ref{ap_contamination}.

The XLFs derived for the $M_r < -18.5$ and $M_r < -19$ selections exhibit very similar shapes, with signs of incompleteness emerging in the lowest-luminosity bin ($L_X \sim 2.5\times10^{42}$ erg s$^{-1}$). For the $M_r < -19.5$ sample, incompleteness appears at slightly higher luminosities, reflecting its intermediate depth between the $M_r < -19$ and $M_r < -20$ selections. Identifying the luminosity at which incompleteness occurs is important for subsequent studies of cluster and group populations, such as cosmological analyses, where incomplete catalogs can introduce potential biases.

At brighter cuts ($M_r < -20$), which also probe a larger comoving volume, the XLF extends to higher luminosities, populating the upper end of the distribution. However, incompleteness becomes significant below $L_X \sim 10^{43}$ erg s$^{-1}$.

For the purity level of 90\%, the XLFs obtained with different magnitude cuts exhibit similar trends to the 80\% purity case, with a small decrease of $\sim7\%$ in the XLF amplitude within the same luminosity range. When considering the purest sample ($P = 95\%$), the incompleteness already becomes evident at $L_X < 1.5\times10^{43}$ erg s$^{-1}$ for all magnitude cuts. This value is obtained after a drop of almost $52\%$ in the XLF values when compared to the same range at $P=90\%$. 

The comparison across purity levels reveals a systematic trend: the most complete samples (${P=80\%}$) show an enhancement of nearly 35\% at the low-luminosity end when compared to ${P=90\%}$, while the purest samples exhibit a deficit of roughly 60\% for high-luminosity clusters. These results align with the previously discussed behavior, in which stricter purity requirements preferentially remove both low-contrast systems and nearby massive clusters affected by projection effects.

\subsection{Comparison with eRedMaPPer cluster and group catalog} \label{sec:red_comp}

\begin{figure}
    \centering
    \includegraphics[width=\linewidth]{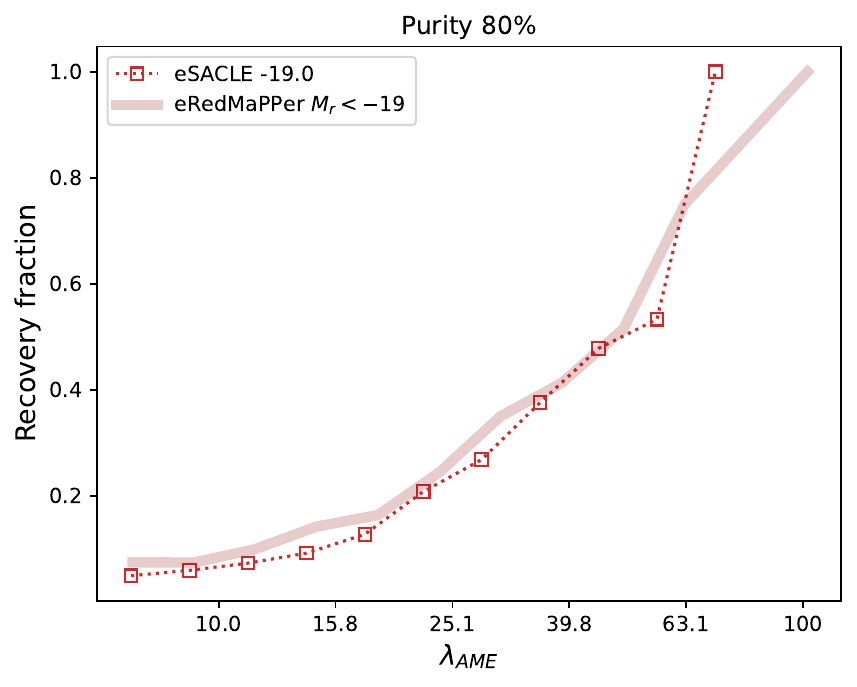}
    \caption{Recovery fraction as a function of the cluster richness, estimated with the AME algorithm. Dark red symbols represent the eRedMaPPer sample over the redshift range $0.08 < z < 0.15$, while red symbols show the eSCALE sample with $M_r < -19$. The similar trends highlight the consistency between the samples and their high purity at increasing richness.}
    \label{fig:recovery_fraction_comp}
\end{figure}

\begin{figure}
    \centering
    \includegraphics[width=\linewidth]{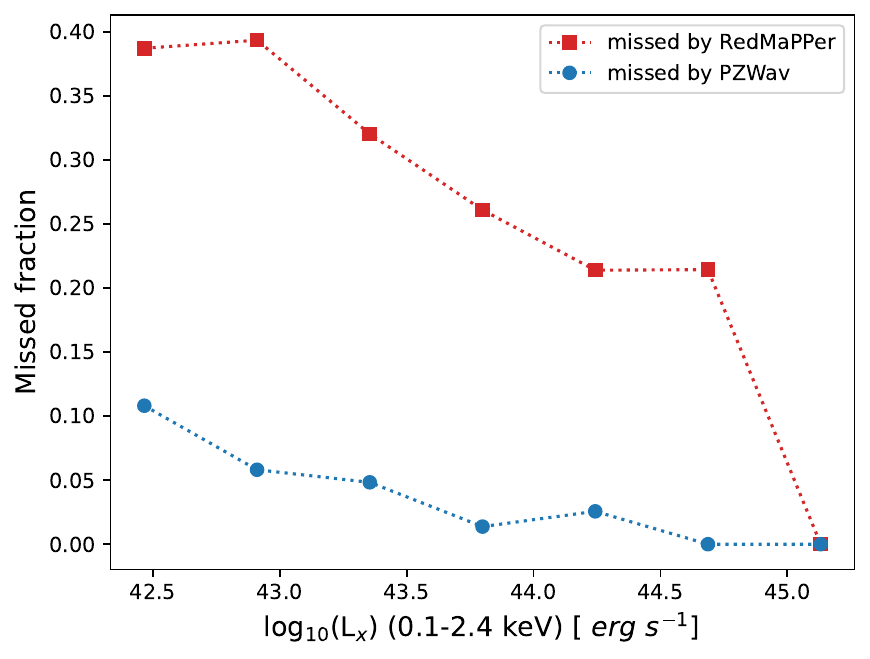}
    \includegraphics[width=\linewidth]{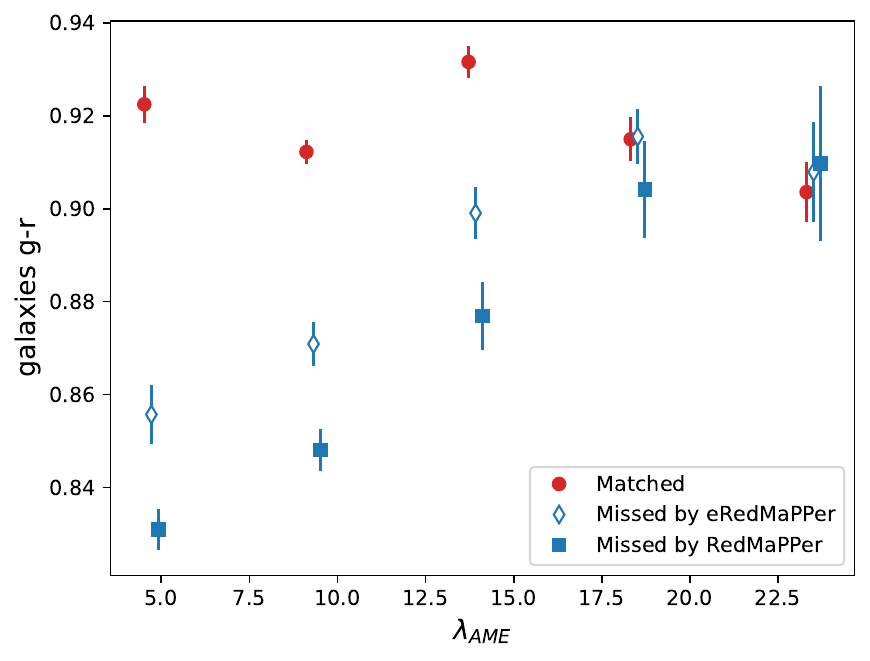}
    \caption{Top panel: Fraction of X-ray sources missed by each catalog relative to their parent optical samples as a function of X-ray luminosity. Filled blue circles show the small fraction of systems not detected by PZWav, while red squares indicate systems not recovered by RedMaPPer.
    Bottom panel: Median color $g-r$ of galaxies in matched and unmatched systems from the RedMaPPer catalog, as a function of the cluster richness. Error bars show the standard error of the median. The missed eSCALE systems are systematically bluer than the matched systems. }
    \label{fig:missed}
\end{figure}

As a cross-validation of our results, we compare our matched catalog with cluster samples produced using analogous X-ray–optical association strategies. 
Specifically, we consider the optical cluster catalog generated by applying the redMaPPer algorithm \citep{Rykoff2014, Kluge2024}, applied to the legacy surveys.
The RedMaPPer catalog comprises 4,029,875 galaxy cluster and group candidates within the redshift range of ${0.012 < z < 0.90}$. Restricting this sample to the S-PLUS footprint and to our redshift limit, $0.08 < z < 0.25$, the number reduces to 21,502 systems. 

We applied the same matching procedure described in Section~\ref{sec:methods} to this subsample, yielding $1,520$ matched systems within the redshift range of $0.08 < z < 0.2$\footnote{The RedMaPPer catalog is all-sky and matched prior to applying the survey footprint. To ensure consistency and limit the data volume, we restrict the sample to $z < 0.2$.}. In matching, we use the membership probabilities as evaluated by the redMaPPer algorithm.  
This sample is referred to as ``eRedMaPPer'' throughout the text. 
While the redMaPPer algorithm employs a different membership assignment scheme -- restricting galaxies to within $0.67R_{200}$ and selecting those brighter than $m < m^* + 1.75$ -- the comparison is performed over the same redshift ranges. We assume this approach probes a broadly similar optical depth, enabling a meaningful comparison between the samples. Production of a random catalog follows the same scheme as for PZWav: adding 10 degrees to the R.A. (modulo 360) of each galaxy and repeating all identification steps. 

Figure~\ref{fig:recovery_fraction_comp} shows the recovery fraction of X-ray sources over the complete optical catalog as a function of the median richness of the catalog for the absolute magnitude cut of $M_r<-19$. As the richness values are not directly comparable due to the different methodologies, we estimate eRedMaPPer richness values using the AME and S-PLUS data, adopting the same absolute magnitude cut. We find a close agreement between the two relations, indicating no potential biases in the recovery rate given the membership selection.

As discussed in the previous subsection, Figure~\ref{fig:XLF} highlights the resulting X-ray luminosity functions, with the eRedMaPPer-based measurements indicated by the shaded red regions. The upper and lower limits correspond to the redshift-limited samples $0.08 < z < 0.13$ ($M_r < -18.5$) and $0.08 < z < 0.2$ ($M_r < -19.5$), respectively. Overall, we find consistent behavior between the eRedMaPPer and eSCALE samples across the luminosity range. 
At the low-luminosity end, however, the eSCALE sample appears more complete, showing higher number counts relative to eRedMaPPer. For this catalog, the incompleteness becomes evident at $L_x=1.17\times10^{43}$ erg s$^{-1}$.

This difference becomes more pronounced in the $\log N$–$\log S$ distribution (Figure~\ref{fig:flux_dist}). Over comparable redshift ranges, the eSCALE counts exceed the eRedMaPPer curve by a factor of 1.5. The black stars indicate the combined contribution of eSCALE and eRedMaPPer systems without an eSCALE counterpart. These additional 232 sources correspond to an increase of $8\%$ relative to the eSCALE $M_r < -20$ sample. When the two samples are combined without applying the redshift cut (small dots), the additional contribution from low-redshift systems fills the high-flux end of the $\log N$–$\log S$ distribution, bringing the total counts into close agreement with literature measurements.

A successful match between the X-ray-selected samples is defined by sharing the same X-ray identification number. For this configuration, the typical redshift offset between the samples is $\Delta z = -0.008 \pm 0.024$, and a projected radius between optical centers of $\Delta R = 3.8 \pm 2.5$ arcminutes. Considering the match between the X-ray sample and pure optical catalogs, a match is defined as a positional coincidence between optical centers within a projected radius of $\Delta R < 5$ arcminutes (corresponding to 0.6 Mpc at $z=0.1$ and 1.5 Mpc at $z=0.2$). Using these criteria, we find $1,254$ matches between eSCALE and eRedMaPPer, including $1,020$ eSCALE sources completely missed by RedMaPPer and $87$ eRedMaPPer sources missed by PZWav (from which $35$ are within masked areas in S-PLUS). This check of failed matches within 5 arcminutes allows to rule out the possibility of a structure being detected by both optical algorithms, but attributed to a different X-ray system as the best identification of an X-ray source.

Although the small fraction of PZWav-missed sources might suggest an issue related to the higher catalog density, this is not the case. The PZWav catalog has a larger number of detections per unit area (19 compared to 9 for RedMaPPer), but these additional sources are likely genuine; if they were dominated by random associations, this would be reflected in the comparison with the random catalogs.
The high completeness of PZWav has been reported in previous studies \citep[typically above 95\% for $M_{200} > 10^{14}$ M$_\odot$;][]{Euclid2019, Werner2022, Thongkham2024}. However, a direct comparison is not straightforward due to differences in calibration, as well as in the redshift and mass ranges considered.

To investigate the differences between the detection methods, the upper panel of Figure~\ref{fig:missed} compares the fraction of X-ray sources missed by each catalog as a function of X-ray luminosity. 
The fraction of eRedMaPPer systems not recovered by PZWav remains low ($\sim5\%$) and approaches zero at the high-luminosity end. In contrast, the fraction of eSCALE systems without a RedMaPPer counterpart reaches $\sim40\%$ at low luminosities and decreases to $23\%$ at $L_X<10^{44.3}$ erg s$^{-1}$, continuing to decline at higher luminosities. 

This behavior reflects the different optical selection strategies adopted by the two cluster finders, being further illustrated in the bottom panel of Figure~\ref{fig:missed}, which compares the median color \texttt{g\_auto-r\_auto} of the galaxies for matched and missed eSCALE systems. The red circles show the median g-r color in the matched X-ray sample, revealing a consistently redder population ($\sim 0.92\%$) across the sampled richness range. The blue squares represent eSCALE systems missed by RedMaPPer, while the open circles indicate systems missed by the matching technique. Both subsamples exhibit a higher fraction of blue population ($\sim 0.85\%$) at low richness, which systematically becomes redder toward higher richness values. At the high-richness end, all samples show similar color values. The larger error bars reflect the smaller number of sources in each bin.  
While the redMaPPer algorithm relies on the red sequence and, therefore, favors evolved galaxy populations, PZWav+AME also incorporates blue cluster members, improving sensitivity to low-mass and less evolved systems.

\begin{figure}
    \centering
    \includegraphics[width=\linewidth]{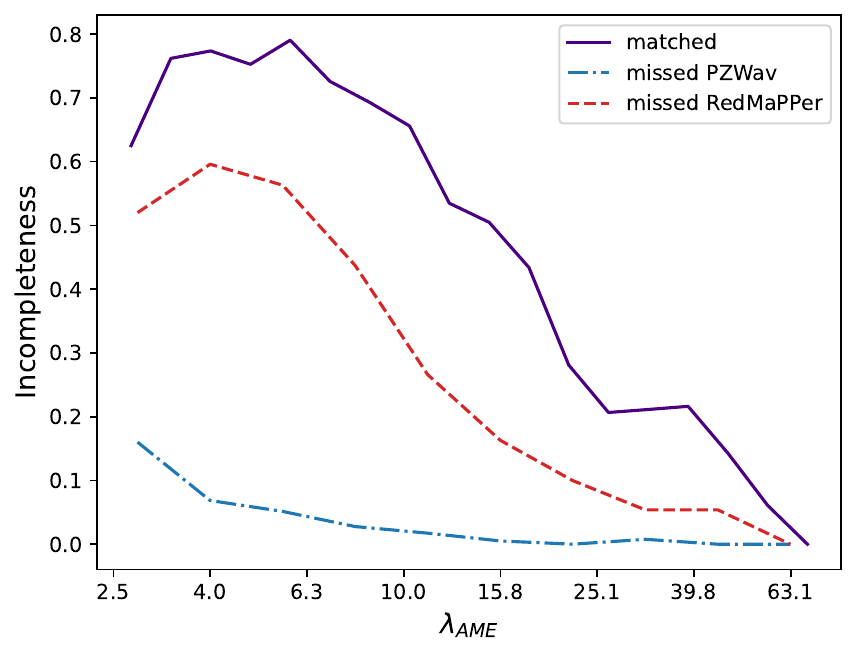}
    \caption{Incompleteness rate as a function of richness for the $M_r < -20$ selection. The matched sample is shown by the purple curve, while systems missed by each algorithm are indicated by PZWav (blue dot-dashed line) and RedMaPPer (red dashed line). The incompleteness of the matched catalog is primarily driven by the RedMaPPer sample at $\lambda_{\mathrm{AME}} < 8$.}
    \label{fig:missed_hist_rich}
\end{figure}

The incompleteness rate distributions as a function of richness (Figure~\ref{fig:missed_hist_rich}) reinforce this interpretation. The matched curve shows that incompleteness becomes significant at $\lambda_{AME} < 8$, driven by the RedMaPPer incompleteness for lower-richness systems.

\section{Conclusions} \label{sec:conclusions}

In this work, we have performed a systematic comparison between optically detected galaxy clusters from the S-PLUS survey and extended X-ray emission traced by eROSITA. By applying a flexible probabilistic matching framework, we explored how optical depth (via absolute-magnitude cuts) and purity requirements influence the construction of matched cluster catalogs and their derived properties. Our analysis quantifies the critical trade-offs between completeness and purity inherent in multi-wavelength cluster identification.

The main results of our study are summarized as follows:

\begin{itemize}
\item The choice of the absolute-magnitude limit critically shapes the recovered cluster population. Fainter cuts ($M_r < -18.5, -19$) enhance sensitivity to low-mass groups, while brighter cuts ($M_r < -19.5, -20$) preferentially select more massive systems and extend the effective redshift range.
\item Purity selection offers a tunable parameter for controlling catalog contamination. Higher purity levels yield cleaner samples but incur a significant reduction in completeness, particularly for low-luminosity and low-mass systems. High-mass clusters are also affected, although to a lesser extent.
\item The $\log N$–$\log S$ distributions confirm that our matched catalogs recover the majority of luminous clusters within the eROSITA footprint while maintaining a higher completeness at low fluxes compared to existing optical cluster catalogs.
\item The X-ray luminosity functions derived from our matched catalogs are in good agreement with previous determinations. Observed deviations at the faint and bright ends are consistently explained by the interplay of incompleteness, purity selection, and survey volume effects.
\item Comparison with the eRedMaPPer cluster catalog reveals consistent scaling trends and significant overlap. Combining both optical selection methods yields a modest improvement in overall detection completeness.
\item Inclusion of blue cluster members through the PZWav+AME approach enhances sensitivity to low-mass and low-luminosity systems, which are underrepresented in red-sequence-selected catalogs such as RedMaPPer. This highlights the importance of multi-color optical selection for a more complete cluster census.

\end{itemize}

In summary, the matching framework presented here provides a robust, reproducible, and adaptable method for associating optical cluster detections with their extended X-ray counterparts. It explicitly accounts for selection effects introduced by optical depth, contamination, and purity requirements. The resulting catalogs offer a valuable resource for cosmological studies and for investigations of galaxy evolution in group and cluster environments. Future work will extend this methodology to deeper eROSITA data releases and wider optical surveys, further refining the synergy between optical and X-ray cluster selections. Within the framework of the SCALE program, these catalogs represent an important step toward building a comprehensive multiwavelength view of galaxy groups and clusters in the S-PLUS footprint.

\section{Data availability}

The eSCALE and eRedMaPPer matched catalogs, and S-PLUS member galaxies presented in this paper are available in electronic form at the CDS via anonymous ftp to cdsarc.u-strasbg.fr (130.79.128.5) or via http://cdsweb.u-strasbg.fr/cgi-bin/qcat?J/A+A/.

\begin{acknowledgements} 
The S-PLUS project, including the T80-South robotic telescope and the S-PLUS scientific survey, was founded as a partnership between the Fundação de Amparo à Pesquisa do Estado de São Paulo (FAPESP), the Observatório Nacional (ON), the Federal University of Sergipe (UFS), and the Federal University of Santa Catarina (UFSC), with important financial and practical contributions from other collaborating institutes in Brazil, Chile (Universidad de La Serena), and Spain (Centro de Estudios de Física del Cosmos de Aragón, CEFCA). We further acknowledge financial support from the São Paulo Research Foundation (FAPESP) grant 2019/263492-3, the Brazilian National Research Council (CNPq), the Coordination for the Improvement of Higher Education Personnel (CAPES), the Carlos Chagas Filho Rio de Janeiro State Research Foundation (FAPERJ), and the Brazilian Innovation Agency (FINEP). The members of the S-PLUS collaboration are grateful for the contributions from CTIO staff in helping in the construction, commissioning and maintenance of the T80-South telescope and camera. We are also indebted to Rene Laporte, INPE, and Keith Taylor for their important contributions to the project. From CEFCA, we thank Antonio Marín-Franch for his invaluable contributions in the early phases of the project, David Cristóbal-Hornillos and his team for their help with the installation of the data reduction package jype version 0.9.9, César Íñiguez for providing 2D measurements of the filter transmissions, and all other staff members for their support with various aspects of the project.
L.D. acknowledges the support from the funding agency FAPESP (grant 2024/03575-9 and 2025/11378-1).
AF acknowledges the financial support from the Visitor and Mobility program of the Finnish Centre for Astronomy with ESO (FINCA). 
C.M.O acknowledges the support from FAPESP (grant 2019/26492-3).
E.V.R.L. acknowledges the support from FAPESP (grant 2024/15229-8).
R.D. gratefully acknowledges support by the ANID BASAL project FB210003.
G.S. acknowledges the support from the  Coordenação de Aperfeiçoamento de Pessoal de Nível Superior - Brasil (CAPES) - Finance Code 001 (88887.177223/2025-00). 

This work is partially based on data from eROSITA, the soft X-ray instrument aboard SRG, a joint Russian-German science mission supported by the Russian Space Agency (Roskosmos), in the interests of the Russian Academy of Sciences represented by its Space Research Institute (IKI), and the Deutsches Zentrum für Luft- und Raumfahrt (DLR). The SRG spacecraft was built by Lavochkin Association (NPOL) and its subcontractors and is operated by NPOL with support from the Max Planck Institute for Extraterrestrial Physics (MPE). The development and construction of the eROSITA X-ray instrument was led by MPE, with contributions from the Dr. Karl Remeis Observatory Bamberg \& ECAP (FAU Erlangen-Nuernberg), the University of Hamburg Observatory, the Leibniz Institute for Astrophysics Potsdam (AIP), and the Institute for Astronomy and Astrophysics of the University of Tübingen, with the support of DLR and the Max Planck Society. 

This paper made use of Astroinspect \cite{astroinspect}.
\end{acknowledgements}

\bibliographystyle{aa}
\bibliography{bib.bib}

\begin{appendix}

\section{Effects of considering source full luminosity} \label{ap_full_lx}

In this appendix, we investigate the impact of the X-ray luminosity definition on our results by considering measurements that include the full X-ray emission of each system, i.e., without removing the central contribution. This central component is known to be enhanced by the presence of active galactic nuclei (AGN) and cool-core clusters, and can therefore bias luminosity-based analyses.

\begin{figure}
    \centering
    \includegraphics[width=\linewidth]{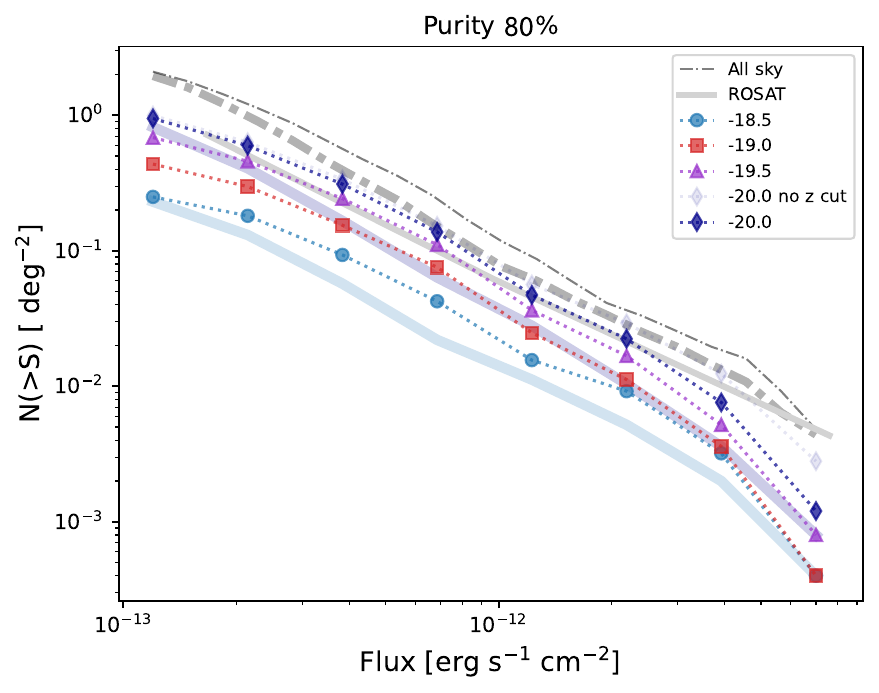}
    \caption{$\log N$–$\log S$ distributions using the full X-ray luminosity of the sources. The $M_r < -20$ sample without redshift cuts is shown for comparison. Soft lines indicate results obtained using the clean flux (Figure~\ref{fig:flux_dist}), with light blue corresponding to $M_r < -18.5$ and dark blue to $M_r < -20$. The dash-dotted line represents the sky coverage contribution. The gray dot-dashed curve shows all sources within the S-PLUS footprint, while the wider dot-dashed curve corresponds to the clean-flux case. When using the full luminosity, the counts exceed the ROSAT All-Sky Survey relation \citep[][gray solid line]{Vikhlinin1998}.} 
    \label{ap_flux_dist_full_Lx}
\end{figure}

For the $\log N$–$\log S$ distribution (Fig.\,\ref{ap_flux_dist_full_Lx}), the differences are most pronounced at the high-flux end. When considering the $M_r < -20$ sample within our adopted redshift range, we already find a good agreement with the literature relation, indicating that the brightest optical selection successfully recovers the luminous cluster population. However, when low-redshift systems are also included, the number counts exceed the reported values in the literature. This highlights that conclusions drawn from $\log N$-$\log S$ analyses are sensitive to the adopted luminosity definition. A fully consistent comparison would require reference samples constructed using a similar treatment of the central X-ray emission, which is currently not available in the literature.

\begin{figure}
    \centering
    \includegraphics[width=\linewidth]{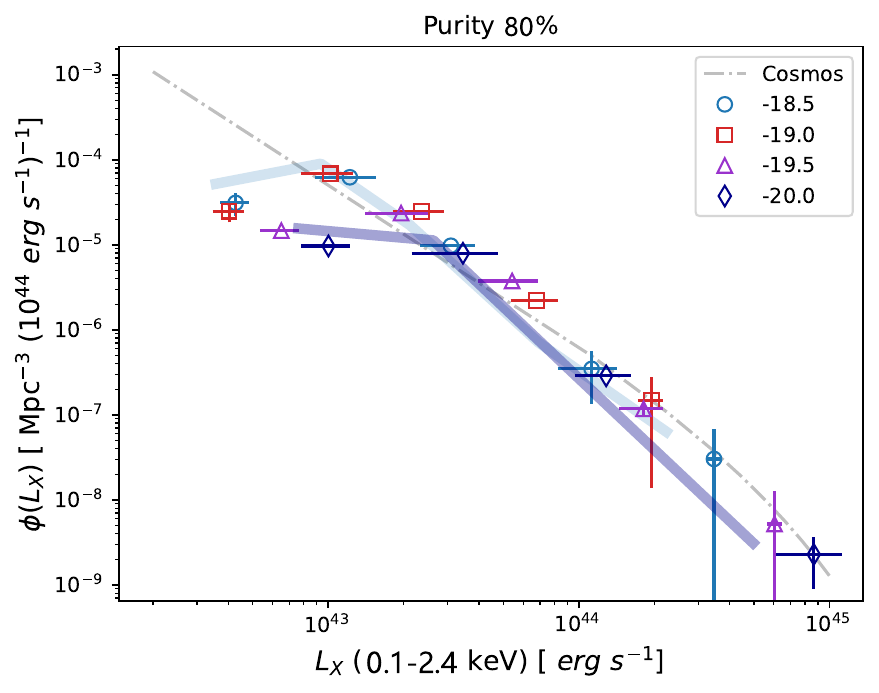}
    \caption{X-ray luminosity functions computed using the full X-ray luminosity of the sources. Soft lines indicate the results obtained with the clean flux (Figure~\ref{fig:XLF}). At the high-luminosity end, the results show improved agreement with COSMOS observations \citep[][gray dot-dashed curve]{Finoguenov2007}, with no clear evidence of evolution, and higher amplitude at $L_x<10^{44}$ erg s$^{-1}$.}
    \label{ap_plot_30xlf_full_Lx}
\end{figure}

For the X-ray luminosity function (Fig.\,\ref{ap_plot_30xlf_full_Lx}), we find an overall good agreement with literature measurements. Nevertheless, we observe a systematic excess of approximately a factor of 1.5 in the number of systems within the luminosity range $L_X \sim 1$–$3 \times 10^{43}$ erg s$^{-1}$. The onset of incompleteness appears consistent between the full and cleaned luminosity estimates, suggesting that the treatment of the central emission does not strongly drive this effect. At the high-luminosity end, we do not observe significant signs of evolution, and all absolute magnitude cuts exhibit similar behavior.

Although the global trends remain broadly consistent, subtle differences between the full and cleaned luminosity definitions are still present, mainly at the bright end. This reinforces that any claim regarding the evolution of the XLF requires a more controlled analysis, ideally adopting homogeneous luminosity definitions across samples.

\section{Random superposition between optical and X-ray sources} \label{ap_contamination}

To quantify the probability of random superpositions between extended X-ray sources and optical systems along the line of sight, we follow an approach inspired by \cite{Klein2018, Klein2019}. In those works, the authors introduced the estimators $P_{\lambda}$ and $P_{\rm S}$, defined as the fraction of random line-of-sight superpositions yielding richness (or signal-to-noise) values lower than that of a given cluster candidate, evaluated within a fixed redshift window around the candidate redshift. These estimators provide an efficient way to identify and remove likely chance alignments between X-ray detections and unrelated optical structures.

A limitation of $P_{\lambda}$ is that it does not explicitly account for the redshift evolution of the richness distribution, which can introduce a redshift-dependent contamination fraction. To mitigate this effect, \cite{Klein2019} introduced alternative contamination estimators that incorporate count-rate and redshift dependencies, allowing the construction of subsamples with approximately redshift-independent contamination.

In our case, we adopt a simplified approach to account for redshift evolution by normalizing the richness by the detection redshift, using the quantity $R_{AME}/z$. For each matched system in the SCALE catalog, we compute the fraction of random line-of-sight objects with $(\lambda/z)_{\rm rand}$ lower than the observed value, within a redshift interval $\Delta z = 0.05$ centered on the cluster redshift. This defines the probability $P$ that a system is not a random superposition. This formulation captures the primary redshift dependence of richness in a direct approach, while preserving the statistical framework introduced by \cite{Klein2019}.

\begin{figure}
    \centering
    \includegraphics[width=\linewidth]{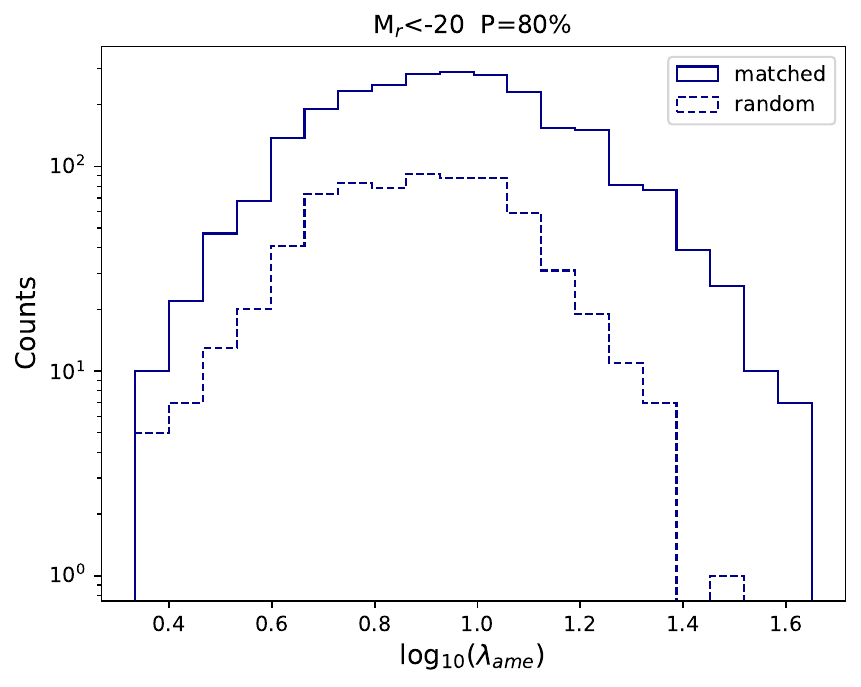}
    \caption{Richness distribution of the SCALE sample with an absolute magnitude cut $M_r<-20$ with a purity of 80\% (filled), compared to the associated random catalog used to estimate the probability of random superpositions (dashed lines). A higher number count is observed at the low richness end.}
    \label{ap_plot_30_20_contamination}
\end{figure}

\begin{figure}
    \centering
    \includegraphics[width=\linewidth]{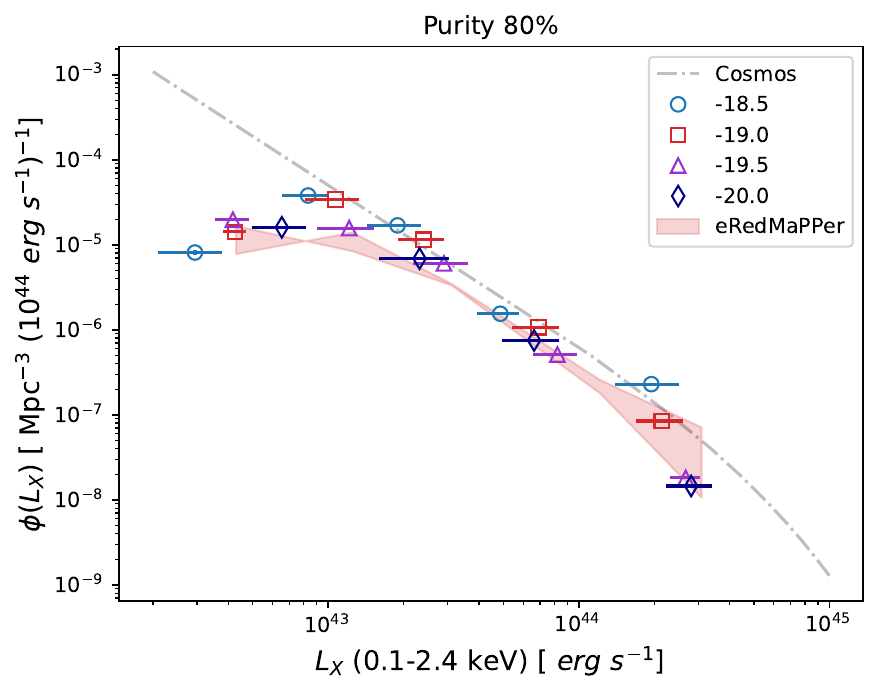}
    \caption{X-ray luminosity functions of the matched clusters after applying a purity threshold of 70\%, corresponding to a maximum 30\% probability of random superposition. Luminosities are corrected for contamination. Different symbols indicate the adopted absolute magnitude cuts applied to the eSCALE sample. The red shaded region shows the eRedMaPPer catalog over $0.08 < z < 0.2$. The results are consistent with COSMOS observations \citep[][gray dot-dashed curve]{Finoguenov2007}, while the eRedMaPPer sample is more affected by the cut at the low-luminosity end.}
    \label{ap_plot_30xlf}
\end{figure}

\begin{figure}
    \centering
    \includegraphics[width=\linewidth]{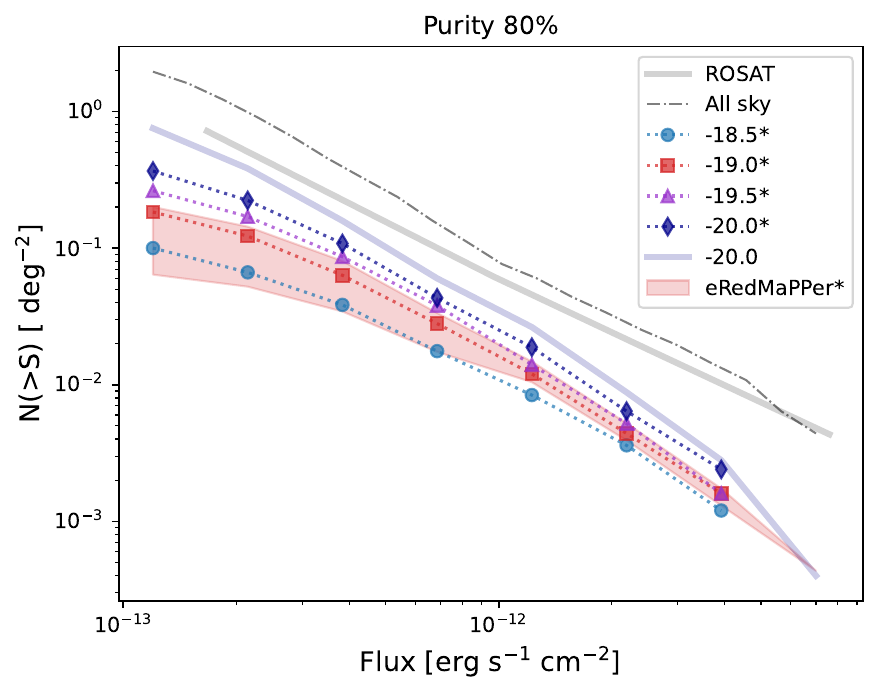}
    \caption{$\log N$–$\log S$ distributions after applying a cut corresponding to a 30\% probability of random superposition (marked by “*” in the legend). The gray solid line shows the ROSAT All-Sky Survey relation \citep{Vikhlinin1998}, and the dash-dotted line indicates the sky coverage.   The cleaning has a stronger impact at the low-flux end compared to the $M_r < -20$ sample shown in Figure~\ref{fig:flux_dist} (soft dark-blue curve).} 
    \label{ap_flux_dist}
\end{figure}

The resulting probability distribution, shown in Fig.\,\ref{ap_plot_30_20_contamination}.
The estimator enables us to identify and remove the lowest-contrast systems through a probabilistic cut.
By selecting systems with $P > 80\%$, corresponding to a probability lower than 30\% of being a random superposition, we construct a cleaned subsample and recompute both the X-ray luminosity function (Fig.\,\ref{ap_plot_30xlf}) and the $\log N$–$\log S$ relation (Fig.\,\ref{ap_flux_dist}). We apply an analogous procedure to the eRedMaPPer catalog to ensure a fair comparison.

The threshold values of $R_{AME}/z$ increase systematically with increasing purity requirement. The values rise by approximately 10\% for the purity level of 90\%, and $\sim20\%$ for P=95\%.

Fig.\,\ref{ap_plot_30xlf} presents the XLF for the $P>80\%$ subsample. The overall behavior remains similar to that shown in the main text, including the relative trends between magnitude cuts. However, at shallower cuts, signs of possible redshift evolution become more apparent, reflected in the reduced number of high-luminosity systems at higher redshifts. Given the sample size, this effect may still be consistent with statistical fluctuations.

Importantly, the previously observed excess at slightly lower luminosities is no longer present, resulting in improved agreement with literature measurements. However, the modeling of the XLF would need to include the richness cut, with the results dependent on the scatter in the richness-luminosity relation below the cut value. The incompleteness limit remains consistent with that derived for the purest sample in the main analysis, becoming evident at $L_X \sim 1.3 \times 10^{43}\,\mathrm{erg\,s^{-1}}$.

For eRedMaPPer, the purity cut has a stronger impact at the low-luminosity end. After applying the $P>80\%$ cut, incompleteness becomes apparent already at $L_X \sim 4 \times 10^{43}\,\mathrm{erg\,s^{-1}}$. Even under this conservative selection, our sample retains superior coverage of low-mass and low-luminosity systems.

Fig.\,\ref{ap_flux_dist} shows the $\log N$–$\log S$ distributions for the cleaned samples. In this case, the impact of the probabilistic cut is more pronounced, producing a noticeable reduction in source counts across nearly the full flux range for both SCALE and eRedMaPPer catalogs. Only the bright end remains close to completeness for $M_{\rm cut} < -20$ when compared to literature determinations.
Although the overall shape of the curves remains consistent with that of the full sample, the low-flux end is more strongly affected, reflecting the preferential removal of lower-contrast systems.

Overall, this analysis demonstrates that our matched catalog remains competitive with existing literature samples even under conservative contamination cuts. The ability to retain a significant population of low-mass and low-luminosity systems highlights the effectiveness of our advanced matching procedure, which incorporates the full galaxy population associated with each structure. This approach reduces the impact of projection effects while preserving sensitivity to less evolved systems.

\end{appendix}

\end{document}